\documentstyle[12pt]{article}
\title{\large\bf HERA high $Q^2$ events as indications of excited \\
leptons with weak isotopic spin 3/2}
\author{\large
B.A.Arbuzov \\
\small\it{Institute for High Energy Physics, Protvino,
Moscow region, 142284, Russia}}
\date{ }
\begin{document}
\newcommand{\bi}{\bibitem}
\newcommand{\be}{\begin{equation}}
\newcommand{\ee}{\end{equation}}
\newcommand{\beq}{\begin{eqnarray}}
\newcommand{\eeq}{\end{eqnarray}}
\maketitle
\bigskip
\begin{quote}
{\small The H1 and ZEUS anomalous events are interpreted as being
due to the production and the decay of excited leptons $E$, which
correspond to spin 1/2 resonances of the first generation lepton
doublet ($\nu_e,\,e$) with W triplet. This assumption is supported by
considering of Bethe-Salpeter equation in the ladder approximation
with anomalous triple gauge boson vertex. The solution with weak
isospin $I = 3/2$ is shown to exist for zero mass state, that means
$M_E$ is small in comparison with $TeV$ mass scale. The coupling
of $E$ with leptons and $W$ is defined by the normalization
condition. Calculation of the $E$ width and the production
cross-sections agrees with HERA data for value of the triple $W$
coupling constant $\lambda \simeq 0.5$. Isotopic relations
for different channels are presented as a tool for checking the
interpretation.}\\
\medskip


\end{quote}
\bigskip

Recent results of H1 \cite{H1} and ZEUS \cite{ZEUS}
collaborations at HERA cause a number of interpretations being
proposed, which mostly deal with leptoquark possibilities. In
the present note I would discuss an excited lepton as a suitable
object, which could manifest itself in data~\cite{H1,ZEUS}. The
excited lepton is understood as a resonant state of the electron
(or the electron neutrino) and $W$. The doublet
$\psi_{eL} = (\nu_{eL},\, e_L)$ and
the $W$ triplet have weak isotopic spin, respectively, $1/2$ and
$1$. Thus weak isotopic spin of their resonance may be either
$3/2$ or $1/2$. We shall see, that from the point of view of both
the theoretical arguments and the experimental data, prescription
$I = 3/2$ is the most favorable. This state $E$ is to have large
mass $M \simeq 200\,GeV$.

Let us start with theoretical arguments. We consider the variant of
the EW theory, in which the symmetry breaking is due to a
self-consistent appearance of additional triple gauge boson vertex
in the region of small momenta~\cite{Arb1,ArbPav}, which can be
described by the following effective term
\be
\Delta\,L\,=\,\frac{\lambda g}{6\,W_W^2}\,\epsilon^{abc}\,
W^a_{\mu\nu} W^b_{\nu\rho} W^c_{\rho\mu}\,.\label{lag}
\ee
Note, that this term is currently considered~\cite{Hag1,Hag2}
among other ones in line of phenomenological analysis of possible
gauge boson interactions. Interaction~(\ref{lag}) defines a vertex
in the momentum space, which according to
the approach acts in the region $p_i^2 < \Lambda^2$, where $p_i$ are
three momenta of the vertex legs and the effective cut-off $\Lambda$
is of the order of magnitude of few $TeV$~\cite{ArbSh,Arb2}. The
vertex has the form
$$
\Gamma^{abc}_{\mu\nu\rho}(p, q, k)\,=\,\epsilon^{abc}\,
\frac{\lambda g}{M_W^2}\,\Gamma_{\mu\nu\rho}(p, q, k)\,;
$$
$$
\Gamma_{\mu\nu\rho}(p, q, k)\,=\,
g_{\mu\nu}(p_\rho (q k) -
q_\rho (p k)) + g_{\nu\rho}(q_\mu (p k) - k_\mu (p q))\,+
$$
\be
+\, g_{\rho\mu}(k_\nu (p q) - p_\nu (q k)) + k_\mu p_\nu q_\rho -
q_\mu k_\nu p_\rho\,.\label{vert}
\ee

Let us now consider the Bethe-Salpeter (BS) equation for the system
consisting of $\Psi_{eL}$ and $W$. Mass parameters $M_W,\,M$
turn out to be much less than both
$\Lambda,\,M_W \sqrt{4 \pi/ \lambda \alpha_w}$, the
last one, as we shall see, is the intrinsic dimensional parameter
of the equation. So it looks reasonable to consider at the
approximation $M_W = M = 0$. In this case we have just $W^0$ as
the neutral component and it does not matter that it consists of
$Z$ and $\gamma$. We consider the BS equation in the ladder
approximation in which the upper vertex in the kernel is just
expression~(\ref{vert}) and the lower one is the usual
$\frac{g}{2}\,\bar \psi_L \gamma_\rho \tau^a \psi_L W^a_\rho$ term.
Let us write down the equation for $I = 3/2$ state
\be
\phi_\mu(p)\,=\,\frac{i\, \lambda g^2}{4 (2\pi)^4 M_W^2}\int
\frac{d^4 q}{(q^2)^2 (p-q)^2}\,\Gamma_{\mu\nu\rho}(p,-q,q-p)\,
\phi_\nu(q)\,\hat q\,\gamma_\rho (1 + \gamma_5)\,;\label{eqMin}
\ee
For $I = 1/2$ state the coefficient afore the integral has extra
factor $(-2)$. Now we decompose BS wave function in terms of Dirac
matrix structures, taking into account conditions of gauge invariance
\be
\phi_\mu(p)\,=\,\sigma_{\mu\nu} p_\nu\,(1 + \gamma_5)\,F_1(p^2)\,+
\,(p^2 \gamma_\mu - \hat p p_\mu)\,(1 + \gamma_5)\,F_2(p^2)\,;
\label{wf}
\ee
After usual trace calculations and the Wick rotation we
obtain identical equations for both functions $F_i\;(i = 1,\,2)$
\be
F_i(p^2)\,=\,\frac{2 \lambda g^2}{3 (2\pi)^4 M_W^2}\,\int\frac
{(p^2 q^2 - (p q)^2)}{p^2 q^2 (p-q)^2}\,F_i(q^2)\,d^4 q\,;
\label{eqEuc}
\ee
where now $p$ and $q$ are Euclidean momenta.
We can obtain an explicit solution of equation~(\ref{eqEuc})
following well-known old method~\cite{ArbFil}. Namely, after
the angular integration we get from~(\ref{eqEuc})
\beq
& &F_i(x)\,=\,\frac{\lambda g^2}{96 \pi^2 M_W^2}\,
\biggl(-\,\frac{1}{x^2}\int_0^x y^2
F_i(y) dy\,+\,\frac{3}{x}\int_0^x y F_i(y) dy\,+\nonumber\\
& &+\,3 \int_x^\infty F_i(y) dy\,-\,x \int_x^\infty
\frac{F_i(y)}{y} dy
\biggr)\,;\label{eqxy}
\eeq
where $x = p^2,\;y = q^2$. Integral equation~(\ref{eqxy}) is
equivalent to the following differential one
\be
\Bigl(x\frac{d}{dx} + 2 \Bigr)\Bigl(x\frac{d}{dx} + 1 \Bigr)
\Bigl(x\frac{d}{dx} \Bigr)\Bigl(x\frac{d}{dx} - 1 \Bigr)\,F_i(x)\,
-\,\frac{\lambda g^2}{16 \pi^2 M_W^2}\,x\,F_i(x)\,=\,0\,;\label{difeq}
\ee
provided integrals in equation~(\ref{eqxy}) converge at infinity and
$F(0) < \infty$. The solution of this boundary problem is expressed
in terms of Meyer functions~\cite{BetErd}. Namely, the solution of
equation~(\ref{difeq}), which satisfies the boundary conditions,
reads
\be
F_i(x)\,=\,C_i\cdot G^{20}_{04}( h x | 1,\,0,\,-1,\,-2 )\,;\qquad
h\,=\,\frac{\lambda g^2}{16 \pi^2 M_W^2}\,.\label{solu}
\ee
The behavior of this function at boundary points is the following
\beq
& &G^{20}_{04}( 0 | 1,\,0,\,-1,\,-2)\,=\,\frac{1}{2}\,;\nonumber\\
& &G^{20}_{04}( x | 1,\,0,\,-1,\,-2)\,\sim\,x^{-7/8}\cdot
\sin(4 x^{1/4} + \delta)\,,\quad x\to\infty\,.\label{asym}
\eeq
From here we immediately see, that the solution exists only for
$\lambda > 0$; this is, so to say, the attraction condition.
For the other sign asymptotic formula~(\ref{asym}) gives
exponential increase at infinity. Hence
for the sign of $\lambda$, which provides the existence of $I = 3/2$
state, a state with $I = 1/2$ does not exist.

Now we proceed to the definition of constants $C_i$. For the purpose
we use the following one-loop normalization condition, where the
outer legs correspond to exited lepton $E$
\be
\frac{i}{(2 \pi)^4} \int\phi_\mu(q,p) \frac{\hat p - \hat q}
{q^2 (p-q)^2}\bar \phi_\mu(q,p)\,d^4 q\,=\,\hat p f_1(p^2) - f_2(p^2)
\,;\label{norm}
\ee
$$
f_1(0)\,=\,1\,;\qquad f_2(0)\,=\,M\,;
$$
where $\bar \phi$ means the Dirac conjugated quantity. In
condition~(\ref{norm})
we need vertex function not only for $p = 0$ as in~(\ref{vert}),
but also for small nonzero $p$. Let us write the expression (k = p-q)
\be
\phi_\mu(q,p)\,=\,\Bigl(\sigma_{\mu\nu} q_\nu F_1(q^2) +
\Bigl((k^2-p^2)\gamma_\mu - (k_\mu+p_\mu)(\hat k - \hat p)\Bigr)
F_2(q^2)\Bigr)\,(1 + \gamma_5)\,;
\label{struc}
\ee
Here we take into account the exact Lorentz-Dirac gauge invariant
structures and assume, that form-factors $F_i$ depend on the
W-boson momentum squared $q^2$. Now from expressions~(\ref{norm}),
(\ref{struc}) we obtain in Euclidean space
\be
\frac{3}{16 \pi^2}\int_0^\infty F_1^2(x) dx\,(1-\gamma_5)\,+\,
\frac{3}{4 \pi^2}\int_0^\infty F_2^2(x) x dx\,(1+\gamma_5)\,=\,1\,;
\label{noreu}
\ee
$$
M\,=\,0\,.
$$
The second line here confirms the consistency of the approach.
Both terms in the first line of normalization relation~(\ref{noreu})
have to be equal to $1/2$. According to asymptotic
formula~(\ref{asym}) the first integral converges, while the
second one diverges. However, we are to remember, that the basic
vertex~(\ref{vert}) has cut-off $\Lambda$. In previous
calculations~\cite{Arb1,ArbSh,Arb2}
we have used Pauli-Villars type of formfactor on each gauge boson
leg. Following this line let us define the following integrals
\beq
& &I_1(\Lambda)\,=\,\int_0^\infty \biggl( G^{20}_{04}( h x | 1,\,0,\,
-1,\,-2)\biggr)^2\,\frac{\Lambda^4}{(x+\Lambda^2)^2}\,dx\,;\label{I12}\\
& &I_2(\Lambda)\,=\,\int_0^\infty \biggl( G^{20}_{04}( h x | 1,\,0,\,
-1,\,-2)\biggr)^2\,\frac{\Lambda^4}{(x+\Lambda^2)^2}\,x\,dx\,.\nonumber
\eeq
Let us take for estimates $\Lambda = 5\,TeV, \lambda = 0.5$, then
a numerical integration gives
\be
I_1(5\,TeV)\,=\,\frac{0.212}{h}\,;\qquad
I_2(5\,TeV)\,=\,\frac{0.233}{h^2}\,.
\label{estint}
\ee
Note that $I_1(\infty) = 1/3 h$.
Constants $C_i$ are calculated with the use of values~(\ref{estint}),
that gives
\beq
& &|C_1|\,=\,\sqrt{\frac{8 \pi^2 h}{3 I_1(\Lambda)}}\,=\,
0.887\,\frac{\sqrt{\lambda}\,g}{M_W}\,; \label{Ci}\\
& & |C_2|\,=\,\sqrt{\frac{2 \pi^2 h^2}{3 I_2(\Lambda)}}\,=\,
0.443\,\frac{\lambda\,\alpha_w}{M_W^2}\,.\nonumber
\eeq
Thus we conclude, that in our approach we show the existence
of $I =3/2$ lepton-$W$ state, which is light in comparison
with few $TeV$ scale.

Now let us proceed to a phenomenology. Relying on the above results,
we assume that there exists spin 1/2 heavy lepton $E$ with mass
$M \simeq 200\,GeV$, corresponding to weak isotopic spin $I = 3/2$,
that means multiplets: ($E^+,\,E^0,\,E^-,\,E^{- -}$) and
($\bar E^{++},\,\bar E^+,\,\bar E^0,\,\bar E^-$). From~(\ref{Ci})
we see, that for an energy scale of few hundreds $GeV$
$C_2$ gives much smaller contribution than $C_1$ and so we may
restrict ourselves by
the following vertex of $(E\,l\,W)$ interaction (mind also
Eqs.~(\ref{asym}), (\ref{struc}))
\be
V_\mu(k)\,=\,i\,\frac{\xi \sqrt{\lambda}\,g}{2 M_W}
<3/2,\,I_z\,| 1,\,I_{W,z};\,
1/2,\,I_{l,z}>\,\sigma_{\mu\nu}\,k_\nu\,(1 +\gamma_5);\label{vertE}
\ee
where usual Clebsch-Gordan coefficient is entering, $I = 1/2$ lepton
means $(\nu_e,\,e)$ pair, $k_\nu$ is the
$W$ momentum and $\xi$ is a dimensionless coefficient of order of
unity ($\xi = 0.887$ from Eq.~(\ref{Ci})). Vertex~(\ref{vertE})
describes an interaction of a left lepton with a right $E$.

Let us begin with $E$ width. It reads
\be
\Gamma_E\,=\,\frac{\xi^2 \pi \lambda \alpha_w (M^2-M_W^2)^2
(3 M^2 - 2 M_W^2)}{8\,M_W^2\,M^3}\,.\label{width}
\ee
The next quantity to calculate is the differential cross-section of
reaction $e^+ + u \rightarrow \bar E^{+ +} + d$
\be
\frac{d\sigma}{dt}\,=\,\frac{4 \pi^2 \xi^2 \lambda \alpha_w^2}
{M_W^2}\,\frac{(s - M^2) (s + t - M^2) (-t)}{s^2 (t-M_W^2)^2}\,;
\label{dif}
\ee
where $M^2 - s < t < 0$. For the total cross-section we have
$$
\sigma(s)\,=\,\int_{m^2-s}^0 \frac{d\sigma}{dt}\,dt\,=
$$
\be
= \frac{4 \pi^2 \xi^2 \lambda \alpha_w^2}{M_W^2}\,
\frac{s-M^2}{s^2}\Bigl( (s+ 2 M_w^2 - M^2)\ln\Bigl(
\frac{s + M_W^2 - M^2}{M_W^2}\Bigr) + 2 M^2 - 2 s \Bigr)\,.
\label{tot}
\ee

For the production of $200\,GeV$ particle in $\sqrt{s} = 300\,GeV$
$e\,P$ collisions,
as it is in experiments~\cite{H1,ZEUS}, large $x$'s are needed.
Bearing in mind, that $u(x)$ in proton is dominant for large $x$,
we use just $u(x)$ to estimate the cross-section in reaction
$e^+ + P \rightarrow \bar E^{+ +} + X$. Thus we have for this process
\be
\sigma(e^+ P \rightarrow \bar E^{++})\,=\,\int_{x_0}^1 u(x) \sigma (x s)
dx\,;\qquad x_0\,=\,\frac{M^2}{s}\,.\label{totP}
\ee
Let us take the simplified expression of structure function $u(x)$
from presentation~\cite{strfun}. From differential
cross-section~(\ref{dif}) we see, that in our case the maximum of
$Q^2\equiv-\,t$ distribution is achieved at
$Q^2\simeq(60\div70\,GeV)^2$. Taking value $Q^2 = (70\,GeV)^2$
and neglecting terms with higher powers of $(1-x)$, we have
$$
u(x)\,=\,3.2\cdot(1 - x)^{3.7}\,.
$$
With this function we obtain the following expression for the total
cross-section at $\sqrt{s}=300\,GeV$, which depends on $m=M/GeV$
\be
\sigma_0(m)\,=\,\xi^2 \lambda\cdot 2.76\cdot 10^{-33}\,I(m)\,cm^2\,;
\label{sigma}
\ee
$$
I(200)\,=\,0.460\cdot 10^{-3}\,;\; I(205)\,=\,0.318\cdot
10^{-3}\,;\; I(210)\,=\,0.217\cdot 10^{-3}\,.
$$

Let us now consider the isotopic relations. Taking into account
values of Clebsch-Gordan coefficients and the usual EW interaction
we have the following values for different cross-sections in terms
of $\sigma_0$ provided only reactions on $u$-quark are taken into
account
\beq
& &e^+ + P \rightarrow \bar E^{+ +} + X\,;\qquad \sigma =
\sigma_0(m)\,;\nonumber\\
& &e^+ + P \rightarrow \bar E^{+} + X\,;\qquad
\sigma = \frac{1}{3} \sigma_0(m)\,;\nonumber\\
& &e^- + P \rightarrow E^0 + X\,;
\qquad \sigma = \frac{1}{3} \sigma_0(m)\,;\label{cross}\\
& &e^- + P \rightarrow E^- + X\,;\qquad
\sigma = \frac{1}{3} \sigma_0(m)\,.\nonumber
\eeq
Now let us consider the necessary decay channels, which we write
below with the corresponding BR in brackets
\beq
& &\bar E^{++}\rightarrow e^+ + W^+\;(100\%)\,;\nonumber\\
& &\bar E^{+}\rightarrow e^+ + W^0\;(66.7\%)\,,\quad
\bar E^+\rightarrow \bar \nu_e + W^+\;(33.3\%)\,;\nonumber\\
& & E^0 \rightarrow e^- + W^+\;(33.3\%)\,,\quad
E^0 \rightarrow \nu_e + W^0\;(66.7\%)\,; \label{decay}\\
& & E^- \rightarrow e^- + W^0\;(66.7\%)\,,\quad
E^- \rightarrow \nu_e + W^-\;(33.3\%)\,.\nonumber
\eeq
From~(\ref{cross}) and~(\ref{decay}) we have total cross-sections for
observable inclusive reactions
\beq
& &\sigma_+^+\equiv \sigma(e^+ + P\rightarrow e^+ + X)\,=\,
\frac{11}{9}\,\sigma_0(m)\,;\nonumber\\
& &\sigma_+^0\equiv \sigma(e^+ + P\rightarrow \bar \nu_e + X)\,=\,
\frac{1}{9}\,\sigma_0(m)\,;\nonumber\\
& &\sigma_-^-\equiv \sigma(e^- + P\rightarrow e^- + X)\,=\,
\frac{1}{3}\,\sigma_0(m)\,; \label{observ}\\
& &\sigma_-^0\equiv \sigma(e^- + P\rightarrow \nu_e + X)\,=\,
\frac{1}{3}\,\sigma_0(m)\,.\nonumber
\eeq
We conclude from~(\ref{observ}), that the positron neutral current
reaction is the most suitable one for studying the effect. This
agrees with data~\cite{H1,ZEUS}.

To compare absolute values of calculated quantities with the data
one needs to define the value of $\lambda$. There are experimental
limitations $|\lambda|\le 0.5$~\cite{lambda1,lambda2}. On the other
hand, theoretical considerations in the framework of EW symmetry
breaking model~\cite{Arb1,ArbPav,ArbSh,Arb2} lead to the conclusion,
that $\lambda$ has to be of order of magnitude of few tenths for
the model being meaningful. So in our approach it seems reasonable
to consider $\lambda$ being close to the experimental limitation.
Bearing in mind also our estimate~(\ref{Ci}) of $\xi$, let us take
\be
\xi^2\,\lambda\,=\,0.4\,.\label{xilam}
\ee
With this value we obtain quite satisfactory total cross-section
for the effect under discussion. For example, for $M=207\,GeV$
and value~(\ref{xilam}) we have
\beq
& &\sigma(e^+ + P \rightarrow e^+ + X)\,=\,0.37\,pb\,;\label{comp}\\
& &\Gamma_E\,=\,14.4\,GeV\,.\nonumber
\eeq
The experimental total cross-section from combined
results~\cite{H1,ZEUS} is $\sigma = (0.32\pm 0.10)\,pb$. The value
of the width also fits data~\cite{H1}.

For possibility $\lambda < 0$, which corresponds to excited
lepton with $I = 1/2$, we obtain the width to be twice
expression~(\ref{width}) and instead of~(\ref{observ}) cross-sections
to be the following
$$
\sigma_+^+\,=\,\frac{1}{9}\,\sigma_0(m)\,,\quad
\sigma_+^0\,=\,\frac{2}{9}\,\sigma_0(m)\,,\quad
\sigma_-^-\,=\,\sigma_0(m)\,,\quad
\sigma_-^0\,=\,\frac{2}{3}\,\sigma_0(m)\,.
$$
These relations do not agree with data~\cite{H1,ZEUS}.
So we have to rest with $I = 3/2$ option.

For checking the mechanism of the effect, relations~(\ref{observ})
can be used. There is also one interesting possibility to check
the interpretation with the use of photons in the final state.
Indeed, $W^0$ contain $\gamma$ with coefficient $\sin^2 \theta_W$.
We have from the previous relations
\beq
& &\sigma_+^{+ \gamma}\equiv \sigma(e^+ + P\rightarrow e^+ + \gamma
+ X)\,=\,0.23\cdot\frac{2}{9}\,\sigma_0(m)\,=\,0.051\,\sigma_0(m)\,;
\nonumber\\
& &\sigma_+^{0 \gamma}\equiv \sigma(e^+ + P\rightarrow \nu_e + \gamma
+ X)\,=\,0\,;\nonumber\\
& &\sigma_-^{- \gamma} \equiv \sigma(e^- + P\rightarrow e^- + \gamma
+ X)\,=\,0.23\cdot\frac{2}{9}\,\sigma_0(m)\,=\,0.051\,\sigma_0(m)\,;
\label{gamma}\\
& &\sigma_-^{0 \gamma}\equiv \sigma(e^- + P\rightarrow \nu_e + \gamma
+ X)\,=\,0.23\cdot\frac{1}{9}\,\sigma_0(m)\,=\,0.026\,\sigma_0(m)\,;
\nonumber
\eeq
Thus 4\% of $e^+\rightarrow e^+$ anomalous high $Q^2$ events and 15\%
of $e^-\rightarrow e^-$ ones contain hard $\gamma$, the mass of the
$(e\,\gamma)$ system being that of $E$. There is also a possibility
to look for leptonic decays of $W$. For example, 9\% of
$e^+\rightarrow e^+$ events contain the second $e^+$, and the same is
valid for an extra $\mu^+$.

Of course, the best way to check the approach as a whole is to
look for anomalous triple $W$ vertex~(\ref{vert}) with
$\lambda\simeq 0.3\div 0.5$. There is a firm hope, that such
check soon can be done at the forthcoming LEP200.

This work is partially supported by the Russian Foundation of Basic
Researches under project 95-02-03704.

\end{document}